\documentclass[pra,twocolumn,aps,10pt]{revtex4-1}

\usepackage{amsmath,amssymb}

\usepackage{graphicx,color}

\usepackage{amsthm}

\usepackage[colorlinks=true]{hyperref}
\usepackage{xfrac}

\usepackage[T1]{fontenc}

\usepackage{bbm}
\usepackage{nicefrac}

\newcommand{\ii}{\mathrm{i}}
\renewcommand{\vec}[1]{\mathbf{#1}}

\let \Re \relax
\DeclareMathOperator{\Re}{Re}
\let \Im \relax
\DeclareMathOperator{\Im}{Im}

\newcommand{\TRST}{\ensuremath{\mathsf{TRS}^t}}
\newcommand{\TRSC}{\ensuremath{\mathsf{TRS}^*}}

\begin{document}
\title{Non-Hermitian Boundary State Engineering \\ in Anomalous Floquet Topological Insulators}

\author{Bastian H{\"o}ckendorf}
\author{Andreas Alvermann}
\email{alvermann@physik.uni-greifswald.de}
\thanks{Corresponding author}
\author{Holger Fehske}
\affiliation{Institut f{\"u}r Physik, Universit{\"a}t Greifswald, Felix-Hausdorff-Str.~6, 17489 Greifswald, Germany}

\begin{abstract}
In Hermitian topological systems, the bulk-boundary correspondence strictly constraints boundary transport to values determined by the topological properties of the bulk.
We demonstrate that this constraint can be lifted in non-Hermitian Floquet insulators. 
Provided that the insulator supports an anomalous topological phase,
non-Hermiticity
 allows us to
 modify the boundary states independently of the bulk,
without sacrificing their topological nature.
We explore the ensuing possibilities
for a Floquet topological insulator with non-Hermitian time-reversal symmetry, where the helical  transport via counterpropagating boundary states can be tailored in ways that 
overcome the constraints imposed by Hermiticity. 
Non-Hermitian boundary state engineering specifically enables the enhancement of boundary transport relative to bulk motion,
helical transport with a preferred direction, and
chiral transport in the same direction on opposite boundaries.
We explain the experimental relevance of our findings for the example of photonic waveguide lattices.
\end{abstract}

\maketitle

Topological states of matter have 
proven to be 
a research topic where fundamental theoretical insights lead almost inevitably to state-of-the-art practical applications~\cite{HasanKane2010, RevModPhys.83.1057, RevModPhys.91.015006}.
A key feature of topological systems is directional transport via chiral boundary states, which is protected by  topological invariants and thus impervious to the imperfections of real-world implementations~\cite{Klitzing, TKNN, Konig2007, Soljacic, stutzer2018}.
In combination with fundamental symmetries~\cite{RevModPhys.88.035005, PhysRevB.78.195125, PhysRevB.96.155118, PhysRevB.96.195303}, especially time-reversal symmetry (TRS) in topological insulators~\cite{KaneMelePRL, Fu}, 
topology even protects bidirectional helical transport via counterpropagating boundary states~\cite{Konig2007, GreifRostock}. 
Such symmetry-protected topological phases emerge in a variety of physical systems~\cite{Haldane, jotzu2014, Slobozhanyuk2016, PhysRevX.5.021031, Yang2015, Yan2015, Lee2018, klembt2018, HockendorfPRB, HAF19},
where they give rise to a wide spectrum of experimentally observable phenomena~\cite{Hsieh, Zhang2009, Rechtsman, PhysRevLett.115.040402, Lustig2019}. 
The recent discovery 
of anomalous topological phases 
in periodically driven (i.e., Floquet) insulators
demonstrates
the singular relevance of
 topological concepts 
 also in 
 systems far from equilibrium~\cite{KitagawaPRB, Rudner, Nathan,  Maczewsky, Mukherjee, HockendorfJPA}.

Only very recently,
the notion of topological 
phases has been extended to non-Hermitian systems~\cite{PhysRevX.8.031079, PhysRevB.99.235112, 2018arXiv181209133K}.
The perception of 
the role of topology in 
this context
is still changing 
rapidly through theoretical investigation and classification~\cite{PhysRevLett.120.146402, PhysRevLett.121.086803, PhysRevLett.121.136802, PhysRevLett.121.026808, PhysRevB.99.201103, Hararieaar4003, PhysRevB.99.241110, 2019arXiv190608782L, Ghatak_2019, 2019arXiv190207217B, PhysRevB.99.121101} as well as  experimental exploration~\cite{bandres2018, PhysRevLett.115.040402, Weimann2016, 2019arXiv190711562H, PhysRevB.99.165148} of non-Hermitian topological phases,
which includes investigation of the interplay of non-Hermiticity and Floquet  dynamics~\cite{PhysRevB.98.205417, 2019arXiv190802066Z}. 
 Intriguingly, topology is expected to protect transport 
 even against
 damping and dissipation~\cite{PhysRevB.99.241110}.

\begin{figure}
\includegraphics[width=\linewidth]{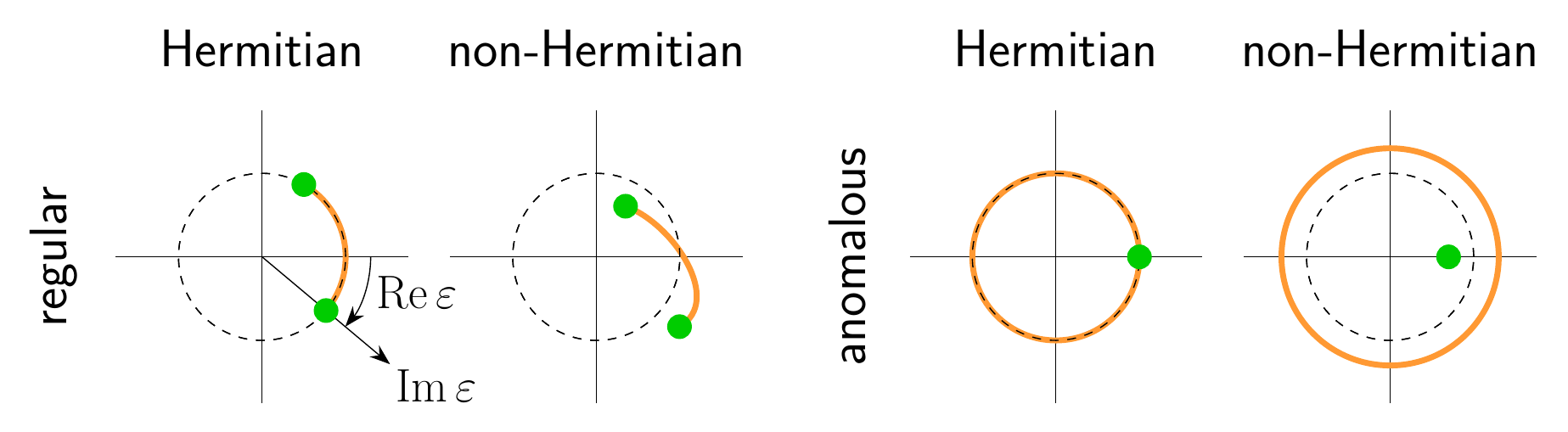}
\caption{Conceptual sketch of the spectrum $\{e^{-\ii \varepsilon}\} \subset \mathbb C$ of the Floquet propagator in the Hermitian and non-Hermitian case, 
with bulk bands (green dots) and boundary states (orange curves) of a regular and anomalous topological phase.
}

\label{fig:concept}
\end{figure}

In the present work we introduce 
the topological concept of \emph{boundary state engineering} (BSE) 
that combines the specific 
aspects of non-Hermitian and anomalous Floquet topological phases,
and has no counterpart in systems with a Hermitian or static Hamiltonian.
The concept underlying BSE 
is illustrated in Figure~\ref{fig:concept}, where we sketch
the spectrum of the Floquet propagator $U \equiv U(T)$, obtained as the solution of the Schr\"odinger equation $\ii \mkern1mu \partial_t  U(t) = H(t) U(t)$
after one period of a time-periodic Hamiltonian $H(t) = H(t+T)$.

For a Hermitian system, with real Floquet quasienergies $\varepsilon$, the spectrum $\{e^{-\ii \varepsilon}\}$ of $U$ lies on the unit circle~\footnote{Note that we measure quasienergies normalized to time step unity.}.
In regular topological phases, as they appear in systems with a static Hamiltonian,
any boundary state, viewed as a continuous curve $k \mapsto e^{-\ii \varepsilon(k)}$ parametrized by momentum
$k$,
connects two different bulk bands.
Anomalous Floquet topological phases, in contrast, possess boundary states that wind around the unit circle~\cite{Rudner}.
Thinking in terms of the quasienergy $\varepsilon$, this possibility 
results from the 
periodicity $\varepsilon \mapsto \varepsilon + 2 \pi$.

In a non-Hermitian system, the spectrum of $U$ can move away from the unit circle.
Regular boundary states  have to remain attached to the bulk bands, since otherwise the continuous dependence on momentum would be violated.
Anomalous boundary states, however, can detach from the bulk bands and 
thus be 
manipulated
independently.
This new freedom is exploited in BSE.

BSE is indeed a topological concept:
Since the propagator $U$ is invertible, its spectrum 
cannot move through the origin, and the winding number is preserved.
Therefore, an anomalous boundary state, which winds around the origin, retains this property during non-Hermitian BSE and remains topologically protected.

\begin{figure}
\includegraphics[scale=0.4]{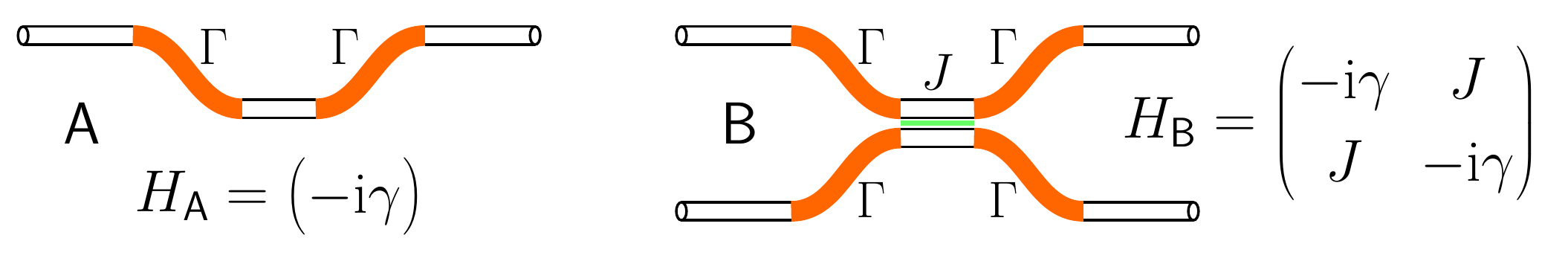}
\caption{Idealized 
Hamiltonians for an uncoupled (\textsf{A}) and two coupled (\textsf{B}) lossy waveguides,
as they might appear in 
the experimental realization of a non-Hermitian driving protocol.
}
\label{fig:waveguide}
\end{figure}

From our discussion it is evident that BSE requires the combination of non-Hermiticity with anomalous Floquet topological phases.
Non-Hermiticity arises naturally in optical settings such as photonic waveguide systems~\cite{PhysRevLett.115.040402},
since coupling involves losses due to the bending of the waveguides (see Fig.~\ref{fig:waveguide}).
The idealized Hamiltonian for the symmetric coupling of two lossy waveguides,
\begin{equation}
H_\mathsf{B} =
 \begin{pmatrix}
 - \ii \gamma & J \\ J & - \ii \gamma
\end{pmatrix} \,, \;
\end{equation}
involves a coupling parameter $J$ and damping $\gamma$.
The associated propagator 
$U_\mathsf{B} = \exp[ -\ii H_\mathsf{B} ]$,
over a time step $\delta t \equiv 1$, is an $\mathrm{SU}(2)$ rotation, modified by the attenuation ($\gamma > 0$) or  amplification ($\gamma<0$) factor $e^{- \gamma}$.
At perfect coupling $J=\pi/2$ we have $U_\mathsf{B} = -\ii {} e^{-\gamma} \hat \sigma_x$,
with the Pauli matrix $\hat \sigma_x$.
Amplitude is swapped between the two coupled sites, but changes as $e^{-\gamma}$.

Regarding the experimental relevance of our theoretical considerations,
 it is useful to allow for a shift $\sigma(t) \in \mathbb C$
of the Hamiltonian,
where we map
$H(t) \mapsto H(t) + \sigma(t)$,
and thus
$U \mapsto \Gamma \, U$
with $\Gamma = \exp(-\ii \int_0^T \sigma(t) \mathrm d t)$.
Through the shift, loss and gain become relative terms,
and weak loss can be interpreted as (pseudo-) gain relative to strong loss.
The physical content of $H(t)$ or $U$ remains unchanged: 
Measuring 
normalized intensities, of the form
 $I(\vec r) = |\psi(\vec r)|^2/\max_{\vec r'}  |\psi(\vec r')|^2$,
 the 
  factor $\Gamma$ in $U$ cancels.

Concerning the second aspect 
of BSE, the anomalous Floquet topological phases,
we resort to the idea of a driving protocol~\cite{Rudner}.
To critically assess the potential of BSE in complex situations, we will use a driving protocol with fermionic TRS that supports counterpropagating boundary states. 
This protocol belongs to a class of universal protocols for symmetry-protected topological phases, which are investigated in Ref.~\cite{HAF19}.

\begin{figure}
\hspace*{\fill}%
\includegraphics[scale=0.39]{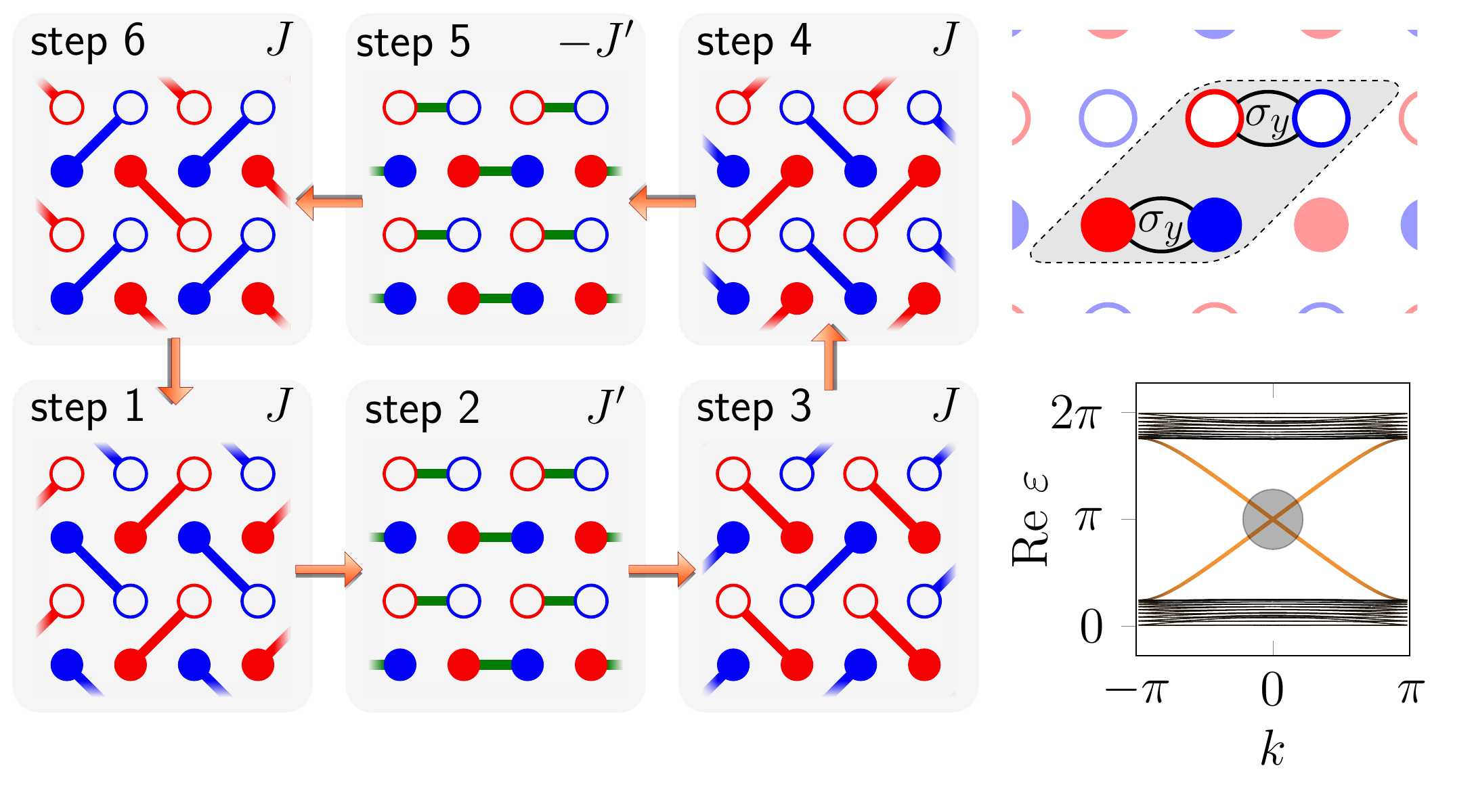}
\hspace*{\fill}%
\caption{Driving protocol for a Floquet insulator with TRS:
Six steps of alternating interactions (left) between neighboring lattice sites (representing, e.g., photonic waveguides) lead to a $\mathbb Z_2$ topological phase with counterpropagating 
boundary states (bottom right), which is protected by fermionic TRS with the symmetry operator $S \equiv \hat\sigma_y \otimes \hat\sigma_y$ (top right).
}
\label{fig:protocol}
\end{figure}

One period of the protocol concatenates $n=6$ steps,
as given in Fig.~\ref{fig:protocol} in a pictorial representation.
The protocol takes place on a (finite or infinite) square lattice, which 
is composed of 
 a ``red'' and ``blue'' sublattice.
Due to fermionic TRS, the Hamiltonian obeys the relation $S H(t) S^{-1} = H(T-t)^*$,
with a unitary symmetry operator $S$ that fulfills $S S^* = -1$~\cite{PhysRevB.78.195125}. 
Here, it is $S = \hat\sigma_y \otimes \hat\sigma_y$ if we identify the ``red'' and ``blue'' sublattice with the up and down component of a pseudo-spin $\nicefrac 12$.

Fermionic TRS is essential for the $\mathbb Z_2$ phases of topological insulators~\cite{KaneMelePRL}.
As can be seen in Fig.~\ref{fig:protocol}, the driving protocol indeed supports a symmetry-protected $\mathbb Z_2$ topological phase, with counterpropagating boundary states whose intersection at momentum $k=0$ is protected by Kramers degeneracy.
This phase is an anomalous Floquet phase, since the boundary states connect the same bulk band at quasienergies separated by $2 \pi$.
This phase has been explored experimentally in Ref.~\cite{GreifRostock}.

In general, the driving protocol supports topological phases on a continuous parameter manifold.
Here, 
we consider a minimal parameter set, with the two parameters $J$ (for diagonal couplings in steps $1, 3, 4, 6$) and $J'$ (for horizontal couplings in step $2, 5$). We use $J = 1.5$, $J'=0.4$ in all plots. 
The minus sign $\pm J'$ between 
steps $2, 5$ shown in Fig.~\ref{fig:protocol} is required for fermionic TRS, 
but negative couplings can be replaced by positive couplings to facilitate the experimental implementation~\cite{GreifRostock}.

Since the $\mathbb Z_2$ phase in Fig.~\ref{fig:protocol} is an anomalous topological phase,
it is a candidate for BSE.
To understand the possibilities arising in this situation
we first address the analytically tractable case of perfect coupling ($J=\pi/2$, $J'=0$),
before returning to general parameters.

\begin{figure}
\includegraphics[width=\linewidth]{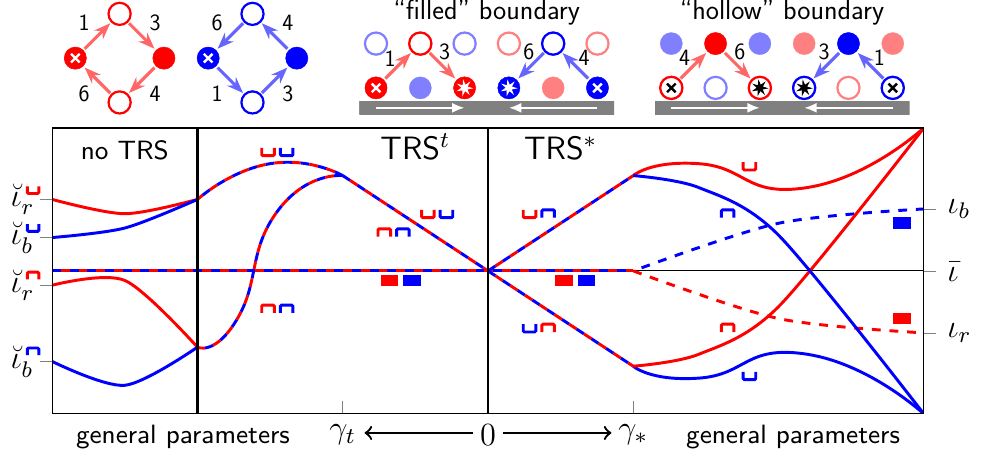}
\caption{Top row: Motion at perfect coupling 
in the bulk (left) 
and at two boundaries  (center and right),
starting from a site marked with a cross ($\times$). 
Lower panel: Schematic plot of the imaginary quasienergy $\iota \equiv \Im \varepsilon$ (relative to $\bar \iota= -2 (\gamma_r + \gamma_b)$) 
for the 
parametrization in Eq.~\eqref{TRSparamIota} and with general parameters.
The plot shows the 
values in the bulk (filled rectangles) and 
on two different boundaries (marked with $\sqcup$, $\sqcap$),
colored according to the starting site as in the top row.
}
\label{fig:imaginary}
\end{figure}

At perfect coupling, the ``red'' and ``blue'' sublattice are decoupled (since $J'=0$).
As depicted in Fig.~\ref{fig:imaginary}, states in the bulk move in a circular clockwise (counterclockwise) orbit on the ``red'' (``blue'') sublattice.
At a boundary, which in Fig.~\ref{fig:imaginary} is oriented horizontally with respect to Fig.~\ref{fig:protocol} and lies at the bottom of the (half-infinite) lattice, states propagate either to the right (red sublattice) or the left (blue sublattice).
The propagation direction 
does not depend on the precise position of the boundary,
but is prescribed 
by the bulk-boundary correspondence.
Here, with counterpropagating boundary states,
transport is protected by topology \emph{and} fermionic TRS.

According to these patterns of motion,
perfect coupling gives rise to a four-fold degenerate dispersionless bulk band at quasienergy $\varepsilon = 0$,
and two counterpropagating boundary states with linear 
dispersion
$\varepsilon_r(k) = \pi + k$, $\varepsilon_b(k) = \pi - k$.
As has been seen in Fig.~\ref{fig:protocol}, these features survive 
qualitatively
for general parameters.

Non-Hermiticity is now introduced into the driving protocol in the following way.
In the bulk, we assume uniform losses (as in the type \textsf{B} configuration in Fig.~\ref{fig:waveguide})
for the diagonal couplings of the red (losses with $\gamma_r$) or blue (losses with $\gamma_b$) sublattice in steps $1, 3, 4, 6$.
The bulk bands thus acquire imaginary quasienergies 
\begin{equation}
\label{IotaBulk}
  \iota_r = -4 \gamma_r \;,
  \quad
   \iota_b = - 4 \gamma_b \;.
\end{equation}
Here and below, a $\iota$ variable denotes the imaginary part $\iota \equiv \Im \varepsilon$ of a 
quasienergy.
 Positive (negative) $\iota$ implies gain (loss) according to 
 $e^{-\ii \varepsilon t} = e^{\iota t} \times e^{-\ii t \Re \varepsilon}$. For the horizontal couplings in step 2, 5 we may assume identical losses, which can be absorbed into the shift 
 of the Hamiltonian and will not be listed explicitly. 

At the boundaries, losses $\breve{\gamma}_r$, $\breve{\gamma}_b$ occur at the isolated sites
that, in the respective step of the protocol, do not couple to other sites (as in the type \textsf{A} configuration in Fig.~\ref{fig:waveguide}).
For example, consider the state starting at a ``filled red'' site in the central top panel in Fig.~\ref{fig:imaginary}.
Through one period of the protocol, this state moves by two sites in steps $1, 3$ (incurring bulk losses $-2 \gamma_r$), and remains at an isolated red site during steps $4, 6$ (incurring boundary losses $-2 \breve \gamma_r$).
Working out the respective patterns for all four states depicted in the top row of Fig.~\ref{fig:imaginary}, we see that the boundary states acquire imaginary quasienergies
\begin{equation}\label{IotaBoundary}
\breve{\iota}_{r}  = \iota_{r}/2 - 2 \breve{\gamma}_{r} \;,
\quad
 \breve{\iota}_{b}  = \iota_{b}/2 - 2 \breve{\gamma}_{b} \;.
\end{equation}

Free choice of the four loss parameters $\gamma_r, \gamma_b, \breve \gamma_r, \breve \gamma_b$
allows for free placement of boundary states relative to the bulk,
as we had anticipated with the concept of BSE introduced in Fig.~\ref{fig:concept}.
In particular, the boundary states are detached from the bulk bands if $\iota_r \ne \breve \iota_r$ or $\iota_b \ne \breve \iota_b$.

The free assignment of losses $\gamma$ (or imaginary parts $\iota$) is not compatible with TRS.
To restore TRS, we can now impose two independent conditions in extension of the Hermitian case, namely
\begin{subequations}
\label{TRS}
\begin{align}
(\mathsf{TRS}^*): \quad S H(t) S^{-1} &= H(T-t)^* + \xi_*(t) \;,\\[0.5ex]
(\mathsf{TRS}^t): \quad S H(t) S^{-1} &= H(T-t)^t + \xi_t(t) \;.
\end{align}
\end{subequations}
In comparison to, e.g., Ref.~\cite{2018arXiv181209133K},
these conditions incorporate the shift of the Hamiltonian
through the functions $\xi_*(t) = \sigma(t) - \sigma(T-t)^*$, $\xi_t(t) = \sigma(t) - \sigma(T-t)$.
For a constant shift $\sigma(t) \equiv \sigma$, we have $\xi_* \equiv \ii \Im \sigma$ and $\xi_t \equiv 0$.

Similar to the coupling parameters $J, J'$, where fermionic TRS requires a minus sign between steps $2, 5$,
non-Hermitian TRS imposes constraints
\begin{subequations}
\label{TRSconstraints}
\begin{align}
(\mathsf{TRS}^*): \quad  & \gamma_r + \gamma_b =  \breve\gamma_b + \breve\gamma_r  \;, \\[0.5ex]
(\mathsf{TRS}^t): \quad & \gamma_r = \gamma_b \;,
 \quad
 \breve\gamma_b =  \breve\gamma_r \;, 
\end{align}
\end{subequations}
on the loss parameters,
which 
are equivalent to
\begin{subequations}
\label{TRSiota}
\begin{align}
(\mathsf{TRS}^*): \quad  & \iota_r + \iota_b =  \breve\iota_b + \breve\iota_r \;, \\[0.5ex]
(\mathsf{TRS}^t): \quad & \iota_r = \iota_b \;,
 \quad
 \breve\iota_b =  \breve\iota_r \;.
\end{align}
\end{subequations}
Note that \TRST{}, but not \TRSC{}, implies equal damping of counterpropagating boundary states.

While in a uniform system the parameters $\gamma_{r,b}$ can be assumed to be 
identical throughout the bulk, 
it is essential for BSE that the losses $\breve \gamma_{r,b}$ may very well depend on the boundary.
For the schematic plot in the lower part of Fig.~\ref{fig:imaginary},
we consider an infinite horizontal strip with different losses 
at the ``top'' ($\breve{\gamma}_r^\sqcap$, $\breve{\gamma}_b^\sqcap$) and ``bottom''
($\breve{\gamma}_r^\sqcup$, $\breve{\gamma}_b^\sqcup$)  boundary.
The central part of this plot, along the $\gamma_*$ and $\gamma_t$ abscissa, uses a parametrization that results in
\begin{subequations}
\label{TRSparamIota}
\begin{align}
(\mathsf{TRS}^*): \quad  &
\iota_{r,b} = -4 \gamma_* \,, \;
\\[0.25ex] 
& \breve{\iota}_{r}^{\,\sqcup} =  \breve{\iota}_{b}^{\,\sqcap} = - 2 \gamma_* \,, \;
\breve{\iota}_{r}^{\,\sqcap} =  \breve{\iota}_{b}^{\,\sqcup} = - 6 \gamma_* \;,
\\[0.5ex]
(\mathsf{TRS}^t): \quad & \iota_{r,b} = -4 \gamma_t \;,
\quad
  \breve{\iota}_{r,b}^{\,\sqcap} =  \breve{\iota}_{r,b}^{\,\sqcup} = -2 \gamma_t
\end{align}
\end{subequations}
for the bulk and boundary states. \TRSC{} or \TRST{} symmetry is preserved for all values of the respective parameter $\gamma_*$ or $\gamma_t$,
and bulk motion is suppressed in favor of boundary transport for $\gamma_{*,t} > 0$. The remaining figures will use
this parametrization of the infinite horizontal strip.

General assignment of losses allows for more complex manipulation of boundary states, as schematically depicted along the ``general parameters'' abscissae. As long as TRS is preserved, the $\iota$ values in Fig.~\ref{fig:imaginary} obey the relations of Eq.~\eqref{TRSiota}. Without TRS, boundary states can be manipulated freely and independently on each boundary, as shown towards the left axis of the plot.

The schematic plot in Fig.~\ref{fig:imaginary}
illustrates 
a few
particularly noteworthy 
features of BSE.
First, without TRS, the damping of bulk and boundary states can be chosen entirely freely.
Second, even with TRS, the damping of boundary states relative to bulk states can be chosen freely,
suppressing or enhancing either boundary transport or bulk motion.
Third, the properties of boundary transport are not dictated by the properties of the bulk,
as a strict bulk-boundary correspondence would demand.
For example at the right axis of the plot, 
boundary transport occurs predominantly via ``red''  states,
opposite to 
the predominant motion in the bulk via ``blue'' states.

\begin{figure}
\hspace*{\fill}%
\includegraphics[width=0.99\linewidth]{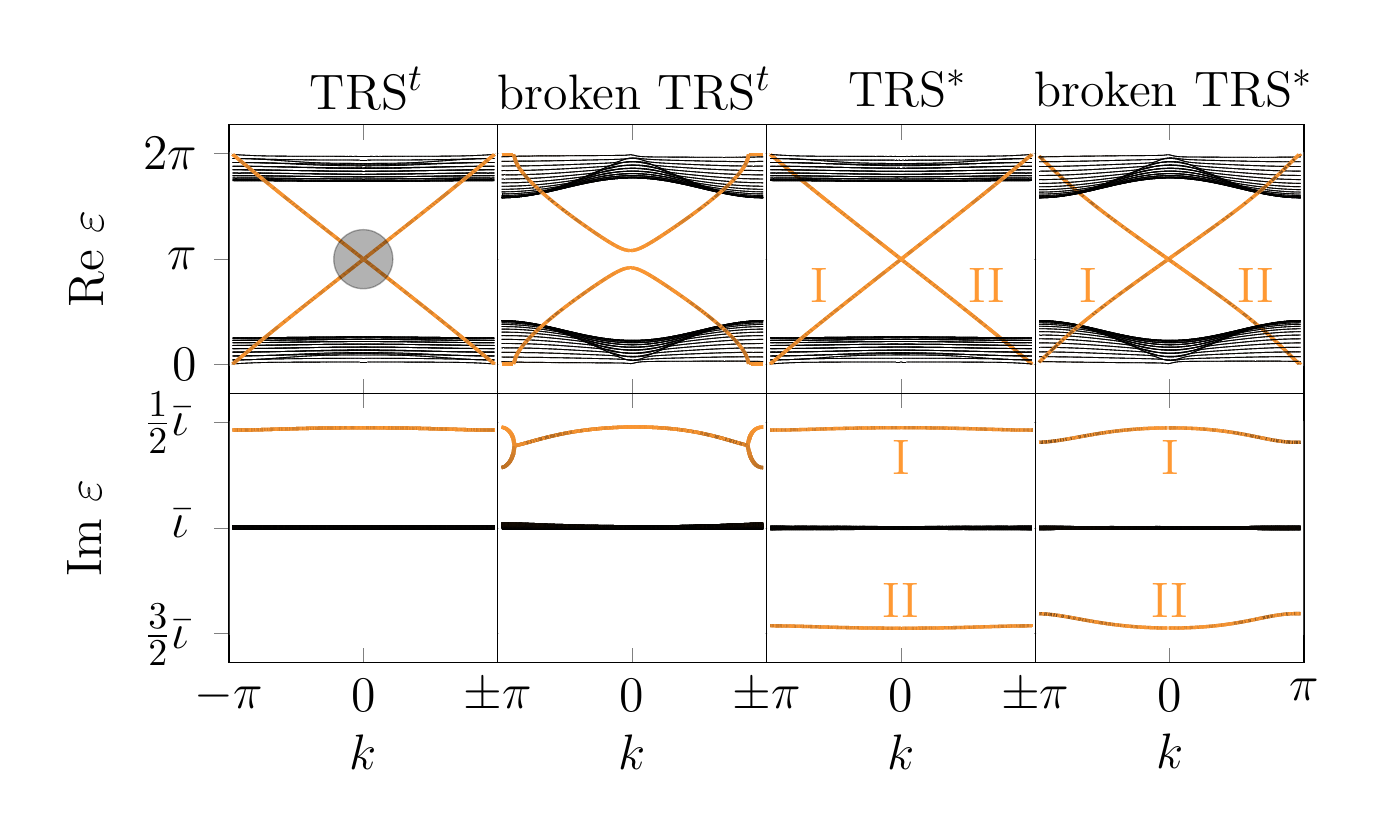}
\hspace*{\fill}%
\caption{Floquet quasienergy $\varepsilon(k)$ as a function of momentum $k$, with and without TRS as indicated,
and non-Hermitian losses $\gamma_{t,*}=1.2$ according to Eq.~\eqref{TRSparamIota}.
Boundary states (on a ``bottom'' boundary) are shown in orange 
(as in Fig.~\ref{fig:protocol}).
The grey circle indicates the Kramers-like degeneracy for \TRST{}.
}
\label{fig:dispersion}
\end{figure}

In the general situation, away from perfect coupling, both the real and imaginary part of the quasienergies depend on momentum.
TRS implies the constraints~\footnote{See the supplemental material for a detailed derivation.}
\begin{subequations}
\label{TRSdispersion}
\begin{align}
 (\mathsf{TRS}^{*,t}): \quad \Re \, \{ \varepsilon(\vec k) \} & = \phantom{-} \Re \, \{ \varepsilon(-\vec k) \} \;, \label{TRSdispersionReal} \\
(\mathsf{TRS}^*): \quad \Im \, \{ \varepsilon(\vec k) \}  &= - \Im \, \{ \varepsilon(- \vec k)  \} + 2 \Im \sigma  \;, \label{TRSdispersionImC} \\
(\mathsf{TRS}^t): \quad  \Im  \, \{\varepsilon(\vec k) \}  &= \phantom{-} \Im  \, \{  \varepsilon(- \vec k) \} 
\label{TRSdispersionImT}
\end{align}
\end{subequations}
on the quasienergy spectrum $\{ \varepsilon(\vec k) \}$ 
at momentum $\vec k$,
which generalize Eq.~\eqref{TRSiota}. Here, Eq.~\eqref{TRSdispersionImC} includes an imaginary shift,
which drops out of Eq.~\eqref{TRSdispersionImT}.

In Fig.~\ref{fig:dispersion}, we observe  the Kramers-like crossing of $\Re \varepsilon(k)$ 
according to Eq.~\eqref{TRSdispersionReal}.
For \TRST{} the two boundary states have to cross at the same $\Im \varepsilon(k)$ and are thus truly degenerate, while for \TRSC{} they are separated by their imaginary part.
This difference suggests that in the first case \TRST{} is required to protect the boundary states,
while in the second case they are robust against breaking of \TRSC{}.
Indeed, if we break TRS by adding detunings~\footnote{For details on TRS breaking and preserving detunings, see the supplemental material.},
we observe an avoided crossing in Fig.~\ref{fig:dispersion} 
only for broken \TRST{} 
but not for broken \TRSC{}.

Evidently, when counterpropagating boundary states have been separated via BSE,
TRS is no longer required for their protection.
However, now a preferred direction of 
transport exists
due to the different damping of the boundary states.
Only if the damping is equal, as in the \TRST{} case in Fig.~\ref{fig:dispersion}, true bidirectional transport without a preferred direction can be observed.
In this scenario, TRS is still required for the protection of transport.

\begin{figure}
\hspace*{\fill}%
\includegraphics[scale=0.42]{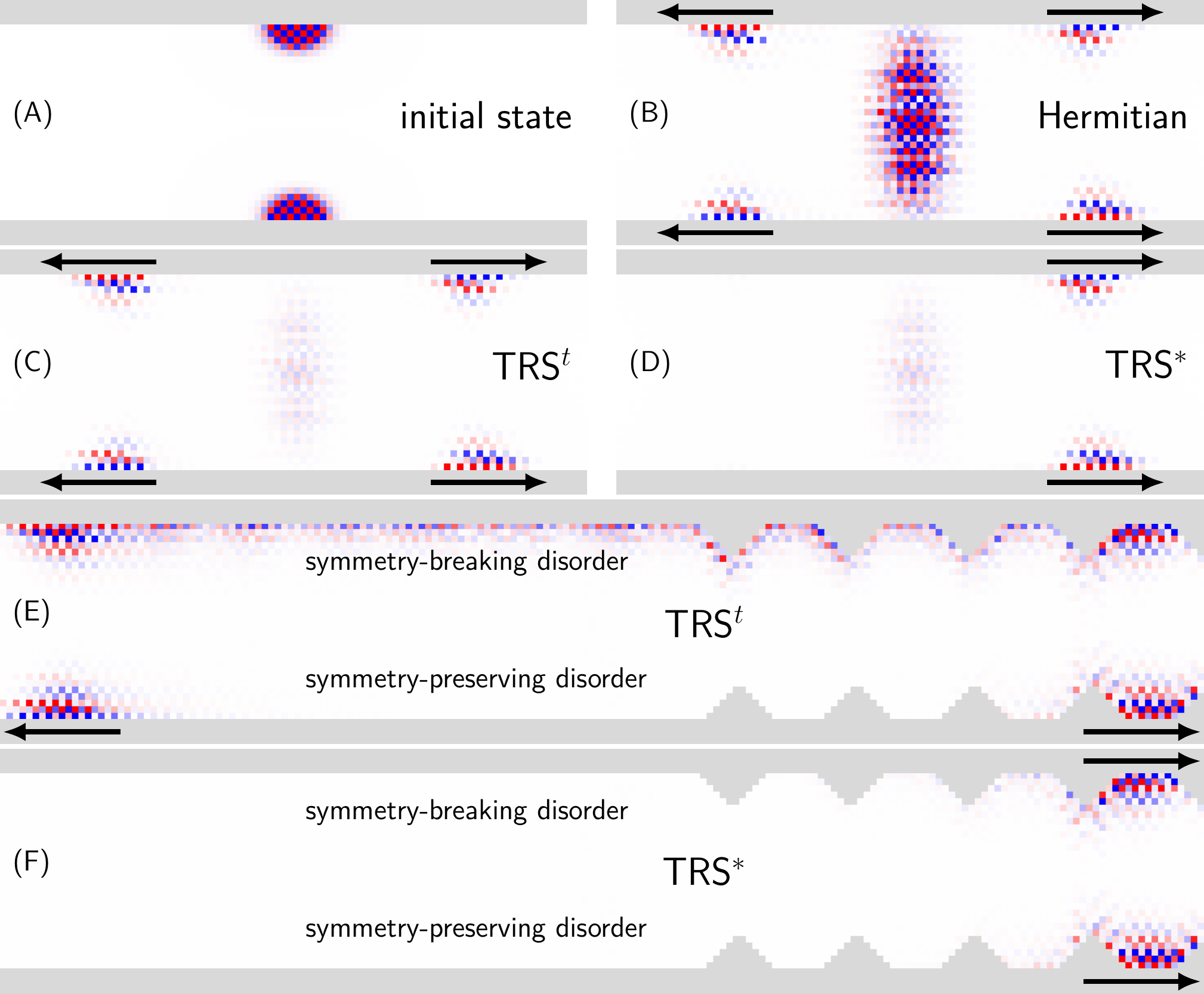}
\hspace*{\fill}%
\caption{Real-space propagation of
an initial state prepared at the boundaries of a semi-infinite strip (height $30$ sites),
in the Hermitian case (panel \textsf{(B)}) or with non-Hermitian
losses $\gamma_{t,*}=0.1$ according to Eq.~\eqref{TRSparamIota} (panels \textsf{(C)}--\textsf{(F)}).
Lightness encodes the wave function intensity $I(\vec r) = |\psi({\vec r})|^2$ at each lattice site,
normalized to the maximum value. 
Panels \textsf{(B)}--\textsf{(D)} show the state after $15$ periods of the driving protocol,
panels \textsf{(E)}, \textsf{(F)} after $45$ periods, with a partly serrated boundary and additional disorder that preserves or breaks TRS.
}
\label{fig:propagation}
\end{figure}

The new freedom introduced by BSE 
is thus two-fold:
It allows us to modify boundary transport relative to the bulk motion,
and
 to selectively modify 
 transport on different boundaries. 
Fig.~\ref{fig:propagation} 
provides a visual demonstration 
of the potential of such modifications. 
The point of reference is the Hermitian case in panel \textsf{(B)}, with bidirectional helical transport via symmetry-protected counterpropagating boundary states (cf. Fig.~\ref{fig:protocol}).
When protected by \TRST{}, the counterpropagating boundary states survive the transition into the non-Hermitian regime,
but now BSE allows us to suppress bulk motion in favor of boundary transport (panel \textsf{(C)}).
In the \TRSC{} case,
we can additionally suppress either one of the two states on each boundary, which we do in such a way that transport on opposite boundaries takes place in the same (and not opposite) direction (panel  \textsf{(D)}).
Such modifications are a unique feature of BSE:
In a Hermitian system,
they are prohibited by the bulk-boundary correspondence, while in a non-Floquet system boundary states remain attached to the bulk bands  (cf. Fig.~\ref{fig:concept}) and thus cannot be selectively amplified or suppressed.

To assess the full extent of topological protection 
in the present 
situation we have to examine the robustness of transport for imperfect boundaries and under the influence of disorder. In panels \textsf{(E)}, \textsf{(F)} in Fig.~\ref{fig:propagation} we use serrated boundaries, and include
 disorder~\footnote{Disorder is both spatial and temporal, the exact form is specified in the supplemental material.
The disorder strength amounts to $0.2$, or $\approx 13\%$ of the coupling $J$.}
that preserves (breaks) TRS in the lower (upper) half of each strip.
In panel \textsf{(E)}, \TRST{} indeed protects the scatter-free bidirectional boundary transport,
while breaking \TRST{} leads to visible back-scattering.
In panel \textsf{(F)},
even disorder that breaks \TRSC{} does not lead to appreciable back-scattering because of the 
suppression 
of one of the two boundary states.
Effectively, this situation realizes chiral transport with a preferred direction, 
which, in contrast to helical bidirectional transport, is protected by topology but does no longer require TRS.

In conclusion, BSE opens up new avenues to control topological transport in non-Hermitian Floquet systems. 
In conjunction with fermionic TRS to protect counterpropagating boundary states,
BSE enables 
the selective enhancement 
of individual topological transport channels,
and thus the effective  manipulation of (bi-)~directional boundary transport.
The potential for applications is immediate, 
and photonic waveguides, which are intrinsically non-Hermitian systems, are a natural platform to explore this potential.
Experiments should also investigate the robustness of the different transport phenomena described here,
thereby extending our analysis of the influence of disorder and symmetry breaking. 
Not least, the surprising new 
possibilities 
of BSE highlight the importance of further theoretical research 
regarding the status of topological invariants 
and  the bulk-boundary correspondence in non-Hermitian Floquet systems.

\end{document}


\title{\textit{Supplemental material for:} \\ Non-Hermitian Boundary State Engineering \\ in Anomalous Floquet Topological Insulators}

\author{Bastian H{\"o}ckendorf}
\author{Andreas Alvermann}
\author{Holger Fehske}
\affiliation{Institut f{\"u}r Physik, Universit{\"a}t Greifswald,  Felix-Hausdorff-Str.~6, 17489 Greifswald,
 Germany}

\begin{abstract}
The supplemental material contains (i) a detailed derivation of Eq.~\eqref{TRSdispersion} in the main text,
(ii) the explicit specification of the non-Hermitian driving protocol,
(iii) the explicit constraints on the protocol parameters from time-reversal symmetry,
(iv) the specification of detunings and disorder to break time-reversal symmetry in Figs.~\ref{fig:dispersion},~\ref{fig:propagation} in the main text,
and (v) three additional examples of non-Hermitian boundary state engineering, including the application to the Kane-Mele model.
\end{abstract}

\maketitle

\manuallabel{fig:concept}{1}
\manuallabel{fig:waveguide}{2}
\manuallabel{fig:protocol}{3}
\manuallabel{fig:imaginary}{4}
\manuallabel{fig:dispersion}{5}
\manuallabel{fig:propagation}{6}

\manuallabel{TRSdispersion}{8}
\manuallabel{TRSconstraints}{5}

\section{Time-reversal symmetry\\ and its consequences}

The standard relation for time-reversal symmetry (TRS) reads $S H S^{-1} = H^* = H^t$ for a static Hamiltonian,
with a unitary operator $S$.
One can distinguish bosonic TRS with $S S^* = 1$ from fermionic TRS with $S S^*=-1$.
The TRS relations generalizes to
\begin{equation}
\label{app:TRSHermitian}
 S  H(t) S^{-1} = H(T-t)^* = H(T-t)^t
\end{equation}
for a Floquet system with period $T$, where $H(t+T)=H(t)$.
This relation generalizes even further for a non-Hermitian Floquet system,
where we have the two separate relations~\cite{2018arXiv181209133K}
\begin{subequations}
\label{app:TRSNoShift}
\begin{align}
(\mathsf{TRS}^*): \quad S H(t) S^{-1} &= H(T-t)^*  \;,\\[0.5ex]
(\mathsf{TRS}^t): \quad S H(t) S^{-1} &= H(T-t)^t  \;.
\end{align}
\end{subequations}
Clearly, both relations agree for a Hermitian Hamiltonian with $H(t)^* = H(t)^t$.

TRS implies relations on the propagator $U(t)$ that lead, eventually, to the relations in Eq.~\eqref{TRSdispersion} in the main text.
To obtain these relations 
it is useful to consider the symmetrized propagator
\begin{equation}
 U_\star(t) = U\Big(\frac{T+t}{2} , \frac{T-t}{2} \Big) \;.
\end{equation}
It is $U_\star(0) = 1$ and $U_\star(T) = U(T)$.
By applying $S \cdots S^{-1}$ on both sides of the equation of motion
\begin{equation}
2 \ii \, \partial_t U_\star(t) = H\Big(\frac{T+t}{2}\Big) \, U_\star(t) \, + \, U_\star(t) \, H\Big(\frac{T-t}{2}\Big) \;,
\end{equation}
replacing terms according to the symmetry relations~\eqref{app:TRSNoShift},
and using the additional equations of motion
\begin{subequations}
\begin{align}
- 2 \ii \, \partial_t U_\star(t)^{-1} &= H\Big(\frac{T-t}{2}\Big) \, U_\star(t)^{-1} \, \notag\\[4pt] & \phantom{=} \quad + \, U_\star(t)^{-1} \, H\Big(\frac{T+t}{2}\Big) \;,
 \\[1ex]
2 \ii \, \partial_t U_\star(t)^{t} &=  U_\star(t)^{t} \, H\Big(\frac{T+t}{2}\Big)^t  \, \notag\\[4pt] & \phantom{=} \quad + \, H\Big(\frac{T-t}{2}\Big)^t \, U_\star(t)^{t} \;,
\end{align}
\end{subequations}
we see that
\begin{subequations}
\begin{align}
(\mathsf{TRS}^*): \quad S U_\star(t) S^{-1} &= (U_\star(t)^{-1})^* \;, \\[0.5ex]
(\mathsf{TRS}^t): \quad S U_\star(t) S^{-1} &= U_\star(t)^{t} \;.
\end{align}
\end{subequations}
Therefore, we have
\begin{subequations}
\label{app:FloquetNoShift}
\begin{align}
(\mathsf{TRS}^*): \quad S U S^{-1} &=  (U^{-1})^*  \;, \\[0.5ex]
(\mathsf{TRS}^t): \quad S U S^{-1} &= U^t 
\end{align}
\end{subequations}
for the Floquet propagator $U \equiv U(T) = U_\star(T)$.

The TRS relations~\eqref{app:TRSNoShift} are somewhat too restrictive for non-Hermitian systems,
where we want to be able to freely interpret the meaning of ``gain'' and ``loss'' in relative terms.
A simple modification suffices to achieve that freedom,
namely, we allow for a (time-dependent) shift $\sigma(t) \in \mathbb C$ of the Hamiltonian
\begin{equation}\label{app:HShift}
 H(t) \mapsto H(t) + \sigma(t)
\end{equation}
and demand that the modified TRS relations are invariant under such shifts.
Replacing $H(t)$ by 
$H(t) + \sigma(t)$
in Eq.~\eqref{app:TRSNoShift}, these modified relations are obtained as
\begin{subequations}
\label{app:TRS}
\begin{align}
(\mathsf{TRS}^*): \quad S H(t) S^{-1} &= H(T-t)^* + \xi_*(t) \;,\\[0.5ex]
(\mathsf{TRS}^t): \quad S H(t) S^{-1} &= H(T-t)^t + \xi_t(t) \;,
\end{align}
\end{subequations}
with arbitrary complex valued functions $\xi_*(t), \xi_t(t) \in \mathbb C$
that fulfill $\xi_*(T-t) = - \xi_*(t)^*$ and $\xi_t(T-t) =  - \xi_t(t)$.
These functions are related to the specific shift $\sigma(t)$ introduced in Eq.~\eqref{app:HShift} through
\begin{subequations}
\begin{align}
(\mathsf{TRS}^*): \quad \xi_*(t) &= \sigma(t) - \sigma(T-t)^* \;,\\[0.5ex]
(\mathsf{TRS}^t): \quad \xi_t(t) &= \sigma(t) - \sigma(T-t) \;.
\end{align}
\end{subequations}

If $H(t)$ is modified by the shift in Eq.~\eqref{app:HShift}, the Floquet propagator is modified as
\begin{equation}
 U \mapsto \Gamma U
\end{equation}
with the scalar factor
\begin{equation}
 \Gamma = \exp\Big( - \ii \int_0^T \sigma(t) \, \mathrm d t \Big) \;.
\end{equation}
With this modification, the TRS relations for the Floquet propagator read
\begin{subequations}
\label{app:TRSFloquet}
\begin{align}
(\mathsf{TRS}^*): \quad S U S^{-1} &=  (\Gamma \Gamma^* U^{*})^{-1} \;, \\[0.5ex]
(\mathsf{TRS}^t): \quad S U S^{-1} &= U^t 
\end{align}
\end{subequations}
in generalization of Eq.~\eqref{app:FloquetNoShift}.
Note that the scalar factor $\Gamma$ drops out of the \TRST{} relation.

Both the complex conjugation in \TRSC{} and the transposition in \TRST{} map momentum $\vec k \mapsto - \vec k$. Therefore, the above relations give
\begin{subequations}
\begin{align}
(\mathsf{TRS}^*): \quad  S U(\vec k, T) S^{-1} &= 
(\Gamma \Gamma^* \, U(-\vec k,T)^*)^{-1} \;, \\
(\mathsf{TRS}^t): \quad S U(\vec k,T) S^{-1} &= 
U(-\vec k,T)^t \;.
\end{align}
\end{subequations}
for the Floquet-Bloch propagator $U(\vec k,T)$ that depends also on momentum $\vec k$.

Thinking in term of the Floquet quasienergy $\varepsilon = \ii \log \lambda$ to eigenvalue $\lambda = e^{-\ii \varepsilon}$ of $U$,
we have that $\Re \varepsilon$ is preserved but $\Im \varepsilon$ changes sign under the mapping $\lambda \mapsto (\lambda^*)^{-1}$. From this, we immediately obtain the relations on the spectrum of the Floquet-Bloch propagator in 
Eq.~\eqref{TRSdispersion} in the main text.

For a driving protocol with discrete steps $k=1, \dots, n$, where the propagator of each step is
$U_k = \exp[-\ii H_k \delta t]$ for the time step $\delta t = T/n$, the above TRS relations can be stated more explicitly.
With a shift
\begin{equation}
 H_k \mapsto H_k + \sigma_k
\end{equation}
in the $k$-th step, and the associated scalar factor $\Gamma_k = e^{-\ii \sigma_k \delta t}$,
we have
\begin{subequations}
\begin{align}
(\mathsf{TRS}^*): \quad  S U_k S^{-1} &= (\Gamma_k \Gamma_{n-k+1}^* \, U_{n-k+1}^*)^{-1} \;, \\
(\mathsf{TRS}^t): \quad S U_k S^{-1} &= 
(\Gamma_{n-k+1}/\Gamma_k) \, U_{n-k+1}^t \;,
\end{align}
\end{subequations}
for the propagators $U_k$ of each step.
If we multiply these equations for all $n$ steps, we see again that the propagator $U \equiv U(T) = U_n \cdots U_1$ of one period of the driving protocol obeys the relations~\eqref{app:TRSFloquet},
now with $\Gamma = \Gamma_1 \cdots \Gamma_n$.

\section{Explicit form of the non-Hermitian driving protocol}

In the general case, the six-step protocol has $6 \times 2 \times 2^2 = 48$ complex parameters.
Hermiticity reduces the number to $24$ real and $12$ complex parameters,
which have been tabulated 
in Ref.~\cite{HAF19} together with the constraints resulting from (fermionic or bosonic) TRS.

For the present study, we choose a restricted set of parameters, with
two coupling parameters ($J$ for diagonal couplings and $J'$ for horizontal couplings),
two  parameters for uniform losses in the bulk ($\gamma_r$ for red and $\gamma_b$ for blue sites)
and individual losses for isolated boundary sites ($\breve \gamma_r$, $\breve \gamma_b$ for each boundary).

Specifically, the bulk Hamiltonian has the following form, using a graphic notation that agrees with Fig.~\ref{fig:protocol} in the main text:

\begin{itemize}
\item[Step] 1: diagonal couplings
\begin{equation}
\label{app:EqProtocolFirst}
\begin{pmatrix}
\textcolor{red}{\bullet} & \swarrow \\
\nearrow & \textcolor{red}{\circ}
\end{pmatrix}_1 =
\begin{pmatrix}
-\ii \gamma_r & J \\[1ex]
 J & - \ii \gamma_r
\end{pmatrix}
\end{equation}

\begin{equation}
\begin{pmatrix}
\textcolor{blue}{\bullet} & \nwarrow \\
\searrow & \textcolor{blue}{\circ}
\end{pmatrix}_1 =
\begin{pmatrix}
-\ii \gamma_b &  J \\[1ex]
 J & - \ii \gamma_b
\end{pmatrix}
\end{equation}

\item[Step]  2: horizontal couplings
\begin{equation}
\begin{pmatrix}
\textcolor{red}{\bullet} & \leftarrow \\
\rightarrow & \textcolor{blue}{\bullet}
\end{pmatrix}_2 =
\begin{pmatrix}
-\ii \gamma_h & J' \\[1ex]
 J' & -\ii \gamma_h
\end{pmatrix}
\end{equation}

\begin{equation}
\begin{pmatrix}
\textcolor{red}{\circ} & \leftarrow \\
\rightarrow & \textcolor{blue}{\circ}
\end{pmatrix}_2 =
\begin{pmatrix}
-\ii \gamma_h & J' \\[1ex]
 J' & -\ii \gamma_h
\end{pmatrix}
\end{equation}

\item[Step] 3: diagonal couplings
\begin{equation}
\begin{pmatrix}
\textcolor{red}{\bullet} & \searrow \\
\nwarrow & \textcolor{red}{\circ}
\end{pmatrix}_3 =
\begin{pmatrix}
-\ii \gamma_r &  J \\[1ex]
  J & - \ii \gamma_r
\end{pmatrix}
\end{equation}

\begin{equation}
\begin{pmatrix}
\textcolor{blue}{\bullet} & \nearrow \\
\swarrow & \textcolor{blue}{\circ}
\end{pmatrix}_3 =
\begin{pmatrix}
-\ii \gamma_b & J \\[1ex]
  J & - \ii \gamma_b
\end{pmatrix}
\end{equation}

\item[Step] 4: diagonal couplings
\begin{equation}
\begin{pmatrix}
\textcolor{red}{\bullet} & \nearrow \\
\swarrow & \textcolor{red}{\circ}
\end{pmatrix}_4 =
\begin{pmatrix}
-\ii \gamma_r &  J \\[1ex]
 J & - \ii \gamma_r
\end{pmatrix}
\end{equation}

\begin{equation}
\begin{pmatrix}
\textcolor{blue}{\bullet} & \searrow \\
\nwarrow & \textcolor{blue}{\circ}
\end{pmatrix}_4 =
\begin{pmatrix}
-\ii \gamma_b & J \\[1ex]
  J & - \ii \gamma_b
\end{pmatrix}
\end{equation}

\item[Step] 5: horizontal couplings
\begin{equation}
\begin{pmatrix}
\textcolor{red}{\bullet} & \leftarrow \\
\rightarrow & \textcolor{blue}{\bullet}
\end{pmatrix}_5 =
\begin{pmatrix}
-\ii \gamma_h & -J' \\[1ex]
 -J' & -\ii \gamma_h
\end{pmatrix}
\end{equation}

\begin{equation}
\begin{pmatrix}
\textcolor{red}{\circ} & \leftarrow \\
\rightarrow & \textcolor{blue}{\circ}
\end{pmatrix}_5 =
\begin{pmatrix}
-\ii \gamma_h & -J' \\[1ex]
 -J' & -\ii \gamma_h
\end{pmatrix}
\end{equation}

\item[Step] 6: diagonal couplings
\begin{equation}
\begin{pmatrix}
\textcolor{red}{\bullet} & \nwarrow \\
\searrow & \textcolor{red}{\circ}
\end{pmatrix}_6 =
\begin{pmatrix}
-\ii \gamma_r &  J \\[1ex]
  J & - \ii \gamma_r
\end{pmatrix}
\end{equation}

\begin{equation}
\label{app:EqProtocolLast}
\begin{pmatrix}
\textcolor{blue}{\bullet} & \swarrow \\
\nearrow & \textcolor{blue}{\circ}
\end{pmatrix}_6 =
\begin{pmatrix}
-\ii \gamma_b &  J \\[1ex]
J & - \ii \gamma_b
\end{pmatrix}
\end{equation}

\end{itemize}

For consistency, we have included losses $\gamma_h$ also for the horizontal steps 2 and 5.
Since these affect all sites equally, they can be absorbed in the shift $H(t) + \sigma(t)$ of the Hamiltonian,
and are thus redundant.

\begin{figure}
 \hspace*{\fill}%
\includegraphics[scale=0.35]{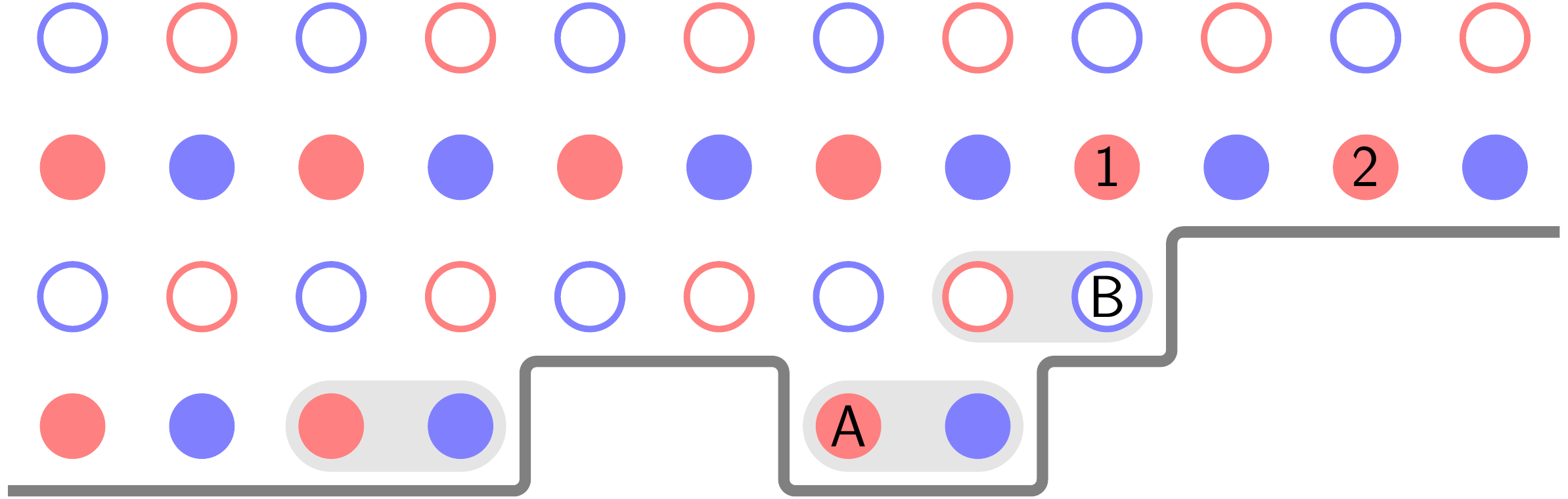}
 \hspace*{\fill}%
 \caption{Sketch of an imperfect ``bottom boundary''.}
 \label{app:fig:imperfect_boundary}
\end{figure}

Isolated boundary sites, i.e., lattice sites that are not coupled in one step of the driving protocol, incur individual losses specified by the parameters $\breve \gamma_r$ and $\breve \gamma_b$ for red and blue sites.
Isolated sites occur only at boundaries, and the set of isolated sites changes during the six steps of the protocol.
In the situation of Fig.~\ref{app:fig:imperfect_boundary}, which shows an imperfect ``bottom'' boundary in generalization of Fig.~\ref{fig:imaginary} in the main text, 
the ``filled red'' site \textsf{A} is isolated in steps $4$ and $6$,
and the ``hollow blue'' site \textsf{B} is isolated only in step $4$.
This statement about lattice sites should not be confused with a statement about states starting at the respective site.
A state starting at site \textsf{A} is a boundary state that (at perfect coupling) moves to site $\mathsf{1}$ and $\mathsf{2}$ in one and two cycles of the driving protocol.
A state starting at site \textsf{B} is a bulk state that (still at perfect coupling) returns to site \textsf{B} in every cycle of the protocol.

Note that with TRS, a boundary has to be compatible with the symmetry operator $S= \hat\sigma_y \otimes \hat\sigma_y$ introduced in Fig.~\ref{fig:protocol} in the main text. For example, if a ``filled red'' site is included also the ``filled blue'' site to the right must be included, as indicated by the grey ovals in Fig.~\ref{app:fig:imperfect_boundary}. This implies that in steps 2 and 5, all boundary sites are coupled through the horizontal couplings $\pm J'$, such that individual losses do not occur.

In principle, the individual losses can differ for each isolated boundary site, restricted only by the corresponding TRS constraints.
In the present study, we only consider the possibility of different losses on different boundaries.

\section{Parameter constraints from time-reversal symmetry}

The constraints on parameter values resulting from TRS can be obtained
along the lines of Ref.~\cite{HAF19}, extended to the non-Hermitian case.
For example, using the graphical notation of the previous section to denote the coupling parameters,
we have with the symmetry operator $S = \hat\sigma_y \otimes \hat\sigma_y$ that
\begin{equation}
S
\begin{pmatrix}
\textcolor{red}{\bullet} & \swarrow \\
\nearrow & \textcolor{red}{\circ}
\end{pmatrix}_1
S^{-1} =
\begin{pmatrix}
\textcolor{red}{\bullet} & \swarrow \\
\nearrow & \textcolor{red}{\circ}
\end{pmatrix}
\overset{\mathrm{TRS}}{=}
\begin{pmatrix}
\textcolor{blue}{\bullet} & \swarrow \\
\nearrow & \textcolor{blue}{\circ}
\end{pmatrix}_6
\end{equation}
relating the parameters of  ``red'' diagonal couplings in step 1 to the ``blue'' diagonal couplings in step 6,
but
\begin{equation}
S
\begin{pmatrix}
\textcolor{red}{\bullet} & \leftarrow \\
\rightarrow & \textcolor{blue}{\bullet}
\end{pmatrix}_2
S^{-1}
=
\begin{pmatrix}
\textcolor{blue}{\bullet} & - (\rightarrow) \\
- (\leftarrow) & \textcolor{red}{\bullet}
\end{pmatrix}
\overset{\mathrm{TRS}}{=}
\begin{pmatrix}
\textcolor{red}{\bullet} & \leftarrow \\
\rightarrow & \textcolor{blue}{\bullet}
\end{pmatrix}_5 \;,
\end{equation}
which explains the minus sign between the horizontal couplings $\pm J'$ in step 2 and step 5.

From these transformations it is straightforward to obtain the constraints required for TRS.
In the general case, say with
\begin{equation}
\begin{pmatrix}
\textcolor{red}{\bullet} & \swarrow \\
\nearrow & \textcolor{red}{\circ}
\end{pmatrix}_1 =
\begin{pmatrix}
A_{r1} & B_{r1} \\[1ex]
 C_{r1} & D_{r1}
\end{pmatrix}
\end{equation}
for the diagonal coupling of red sites in step $1$ and
\begin{equation}
\begin{pmatrix}
\textcolor{blue}{\bullet} & \swarrow \\
\nearrow & \textcolor{blue}{\circ}
\end{pmatrix}_6 =
\begin{pmatrix}
A_{b6} & B_{b6} \\[1ex]
 C_{b6} & D_{b6}
\end{pmatrix}
\end{equation}
for the diagonal coupling of blue sites in step $6$, the TRS constraints on the eight parameters $A_{r1}, \dots, D_{b6}$ are
\begin{subequations}
\begin{align}
(\mathsf{TRS}^*): \quad A_{r1} &= A_{b6}^*+\sigma_1-\sigma_6^* \,, \;  B_{r1} = B_{b6}^* \, ,\\
C_{r1} &=C_{b6}^* \;, D_{r1} =D_{b6}^*+\sigma_1-\sigma_6^* \, ,\\
(\mathsf{TRS}^t): \quad A_{r1} &= A_{b6}+\sigma_1-\sigma_6 \,, \;  B_{r1} = C_{b6} \, ,\\
C_{r1} &=B_{b6} \;, D_{r1} =D_{b6}+\sigma_1-\sigma_6 \, .
 \end{align}
 \end{subequations}
  On the other hand, with 
 \begin{equation}
\begin{pmatrix}
\textcolor{red}{\bullet} & \leftarrow \\
\rightarrow & \textcolor{blue}{\bullet}
\end{pmatrix}_2
=
\begin{pmatrix}
A_{f2} & B_{f2} \\ C_{f2} & D_{f2} 
\end{pmatrix}
 \end{equation}
 for the horizontal coupling of the filled sites in step $2$ and 
  \begin{equation}
\begin{pmatrix}
\textcolor{red}{\bullet} & \leftarrow \\
\rightarrow & \textcolor{blue}{\bullet}
\end{pmatrix}_5
=
\begin{pmatrix}
A_{f5} & B_{f5} \\ C_{f5} & D_{f5} 
\end{pmatrix}
 \end{equation}
 for the horizontal coupling of the filled sites in step $5$, the TRS constraints are
  \begin{subequations}
\begin{align}
(\mathsf{TRS}^*): \quad A_{f2} &= A_{f5}^*+\sigma_2-\sigma_5^* \,, \;  B_{f2} =- C_{f5}^* \, , \\
C_{f2} &=-B_{f6}^* \;, D_{f2} =D_{f5}^*+\sigma_2-\sigma_5^* \, ,\\
(\mathsf{TRS}^t): \quad A_{f2} &= A_{f5}+\sigma_2-\sigma_5 \,, \;  B_{f2} = -B_{f5}\, , \\
C_{f2} &=-C_{f5} \;, D_{f2} =D_{f5}+\sigma_2-\sigma_5\, .
 \end{align}
 \end{subequations}

Analogous constraints are obtained for all other parameters.
In total, TRS introduces $24$ constraints on the $48$ complex parameters of the general protocol.
For the Hermitian protocol, this number reduces to exactly the $12+6$ constraints for the $24+12$ real and complex parameters that have been listed in Ref.~\cite{HAF19}.

For the restricted set of parameters in Eqs.~\eqref{app:EqProtocolFirst}--\eqref{app:EqProtocolLast} used in the present study, where $A_{r1} = - \ii \gamma_r$, $B_{r1} = J$ etc.,
we immediately identify the TRS constraints in Eq.~\eqref{TRSconstraints} in the main text.

\section{Detunings and disorder}

With the minimal set of parameters used for the non-Hermitian driving protocol in the present study, diagonal (i.e., on-site) terms in the Hamiltonian arise from losses and are purely imaginary, e.g., $-\ii \gamma_r, -\ii \gamma_b$
as listed in Eqs.~\eqref{app:EqProtocolFirst}--\eqref{app:EqProtocolLast}.
For Figs.~\ref{fig:dispersion},~\ref{fig:propagation} we add real diagonal terms to the Hamiltonian that account for local potentials or fields.
In a photonic waveguide system, these terms account for detunings that arise from variations in the optical path length of the different waveguides.

In the translational invariant situation, we have four parameters $\Delta_k, \dots, \Delta'''_k \in \mathbb R$ in each step $k=1, \dots, 6$ of the driving protocol,
corresponding to the four different types of sites.
Graphically, we may write the detuning term as
\begin{equation}
\begin{pmatrix}
& \textcolor{red}{\circ} & \textcolor{blue}{\circ} \\
\textcolor{red}{\bullet} & \textcolor{blue}{\bullet} &
\end{pmatrix}_k =
\begin{pmatrix}
& \Delta'_k & \Delta'''_k \\[0.5ex]
\Delta_k & \Delta''_k
\end{pmatrix} \;.
\end{equation}

As before, constraints are required to preserve TRS.
Here, 
the symmetry operator $S$ gives (cf. Fig.~\ref{fig:protocol})
\begin{equation}\label{DisorderConstraints}
S
\begin{pmatrix}
& \textcolor{red}{\circ} & \textcolor{blue}{\circ} \\
\textcolor{red}{\bullet} & \textcolor{blue}{\bullet} &
\end{pmatrix}_k 
S^{-1} =
\begin{pmatrix}
& \textcolor{blue}{\circ} & \textcolor{red}{\circ} \\
\textcolor{blue}{\bullet} & \textcolor{red}{\bullet} &
\end{pmatrix}
\overset{\mathrm{TRS}}{=}
\begin{pmatrix}
& \textcolor{red}{\circ} & \textcolor{blue}{\circ} \\
\textcolor{red}{\bullet} & \textcolor{blue}{\bullet} &
\end{pmatrix}_{n-k+1} \;,
\end{equation}
relating the detunings in steps $1\!\leftrightarrow\!6$,
$2\!\leftrightarrow\!5$, and $3\!\leftrightarrow\!4$.
Since the detunings are real, \TRSC{} and \TRST{} give the same constraints.

For constant detunings $\Delta_k \equiv \Delta$ etc., TRS requires $\Delta=\Delta''$ and $\Delta'=\Delta'''$. In other words, TRS preserving detunings have the form
\begin{equation}
\begin{pmatrix}
& \textcolor{red}{\circ} & \textcolor{blue}{\circ} \\
\textcolor{red}{\bullet} & \textcolor{blue}{\bullet} &
\end{pmatrix} =
\begin{pmatrix}
& \Delta' & \Delta' \\[0.5ex]
\Delta & \Delta
\end{pmatrix} \;,
\end{equation}
with two remaining parameters $\Delta, \Delta' \in \mathbb R$.
In Fig.~\ref{fig:dispersion} in the main text, constant detunings
\begin{equation}
\begin{pmatrix}
& \textcolor{red}{\circ} & \textcolor{blue}{\circ} \\
\textcolor{red}{\bullet} & \textcolor{blue}{\bullet} &
\end{pmatrix}_k \equiv
\begin{pmatrix}
& - \Delta & \Delta \\[0.5ex]
\Delta & - \Delta
\end{pmatrix} 
\end{equation}
with $\Delta=0.5$  are used to break \TRST{} or \TRSC{}.

For the real-space propagation in panels \textsf{(E)}, \textsf{(F)} in Fig.~\ref{fig:propagation} in the main text,
spatial and temporal disorder is included through randomly fluctuating detunings.
For each site $\vec r$ and step $k$ the respective detuning parameter $\Delta^{\vec r}_k$ is independently drawn from a uniform probability distribution in the interval $[-\delta,\delta]$, here with $\delta = 0.2$.
This gives the TRS-breaking disorder used in the upper half of panels \textsf{(E)}, \textsf{(F)}.

To preserve TRS, we enforce the constraints of Eq.~\eqref{DisorderConstraints} locally,
for each pair of sites
$\left(\begin{smallmatrix} \displaystyle\textcolor{red}{\circ} &  \displaystyle\textcolor{blue}{\circ} \end{smallmatrix}\right)$ or $\left(\begin{smallmatrix}  \displaystyle\textcolor{red}{\bullet} &  \displaystyle\textcolor{blue}{\bullet} \end{smallmatrix}\right)$ that is mapped onto itself by the symmetry operator $S$,
and for each period of the driving protocol.
Apart from these constraints, the detunings still fluctuate randomly in space and time.
In particular, they change with each step and period of the driving protocol.
This gives the TRS-preserving disorder used in the lower half of panels \textsf{(E)}, \textsf{(F)}.

\section{Additional examples for non-Hermitian boundary state engineering}

To demonstrate the concept of non-Hermitian boundary state engineering (BSE) in the main text, we choose the TR symmetric driving protocol because it combines experimental relevance with a number of novel transport scenarios.
Being a general concept, BSE applies to any system that supports an anomalous Floquet topological phase. 
Here, we demonstrate the general applicability of BSE with three additional examples.

\subsection{Driving protocols without fermionic time-reversal symmetry}

\begin{figure}
\includegraphics[width=0.4\columnwidth]{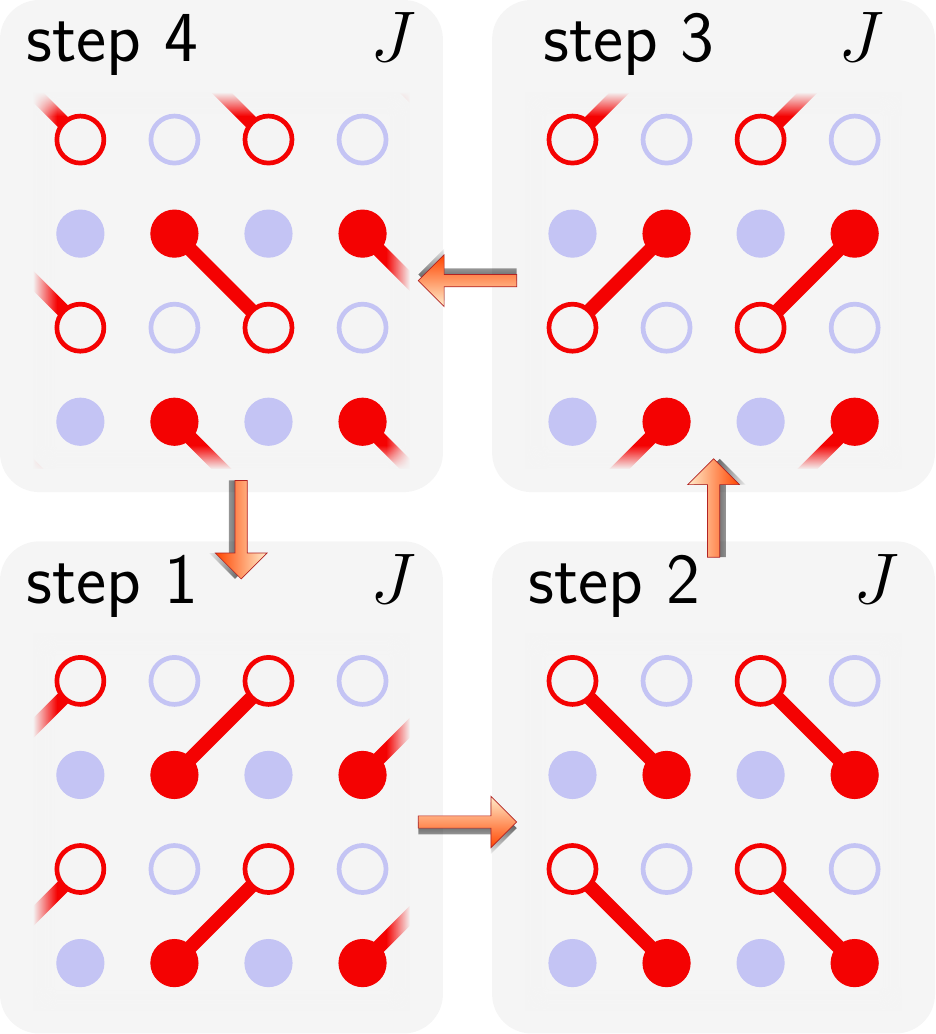}
\caption{Sketch of the four central steps of the driving protocol from Ref.~\cite{Rudner}, depicted in a way that highlights the connection to the TR symmetric driving protocol in the main text~\cite{HAF19}.}
\label{app:fig:Prot1}
\end{figure}

\begin{figure}
\includegraphics[width=\columnwidth]{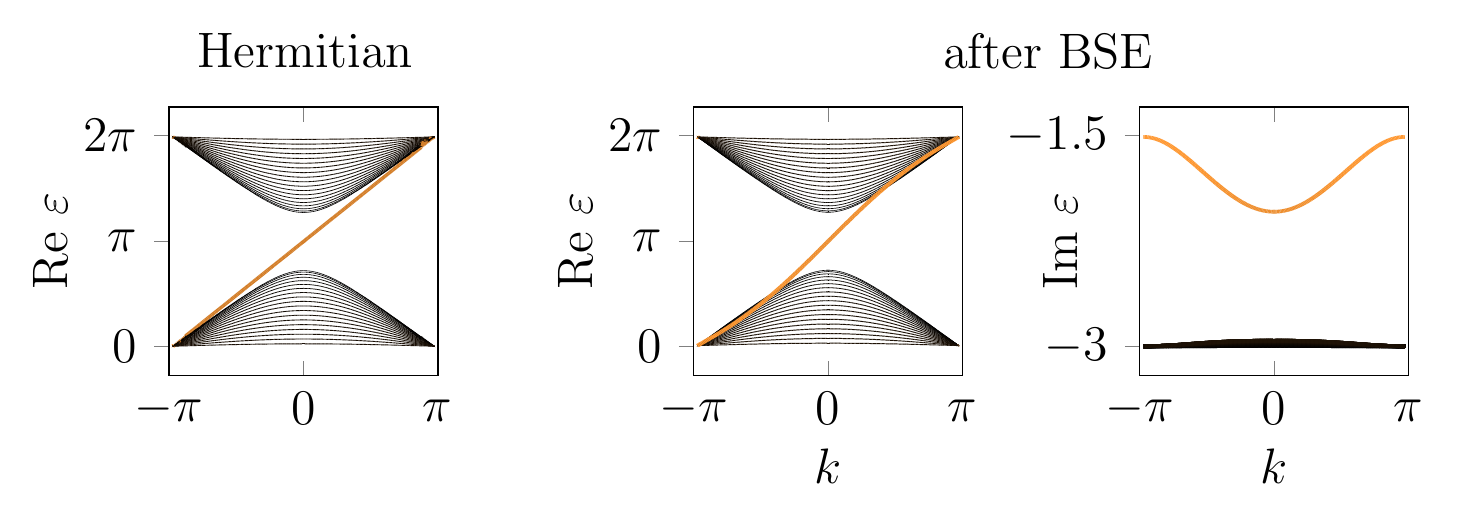}
\caption{
Left panel: Anomalous Floquet topological phase in the driving protocol from Ref.~\cite{Rudner} (with coupling $J=1$).
Central and right panel: 
Bulk and boundary states after BSE (with bulk losses $\gamma = 0.75$).
}
\label{app:fig:Prot2}
\end{figure}

The first example is the driving protocol from Ref.~\cite{Rudner}.
As the sketch in Fig.~\ref{app:fig:Prot1} illustrates, the TR symmetric protocol in the main text essentially comprises two interwoven copies of this protocol, with additional provisions to guarantee TRS and allow for coupling between the ``red'' and ``blue'' sublattice (for details of the construction, see Ref.~\cite{HAF19}).
The present protocol, which by itself does not possess (fermionic) TRS,
features an anomalous Floquet topological phase
with a single chiral boundary state (left panel in Fig.~\ref{app:fig:Prot2}),
instead of the counterpropagating states of the TR symmetric protocol (Fig.~\ref{fig:protocol} in the main text).

To apply BSE to this protocol, we simply introduce uniform losses in the bulk ($\gamma$), excluding the boundary sites
(of course, more complex variants are possible). 
In accordance with the BSE concept, we can thus manipulate the imaginary quasienergies of the bulk and boundary states relative to each other,  and let the boundary states detach from the bulk bands (central and right panel in Fig.~\ref{app:fig:Prot2}).
Here, however, with only a single chiral state per boundary, the potential of BSE is more limited in comparison to what is demonstrated in the main text: Still, boundary transport can be amplified or suppressed relative to bulk motion
(Fig.~\ref{app:fig:Prot2} shows amplification),
but manipulation of the direction of boundary transport as in Fig.~\ref{fig:propagation} in the main text is no longer possible.
This explains why we choose the more complicated TR symmetric driving protocol as our primary example.

\begin{figure}
\includegraphics[width=0.5\columnwidth]{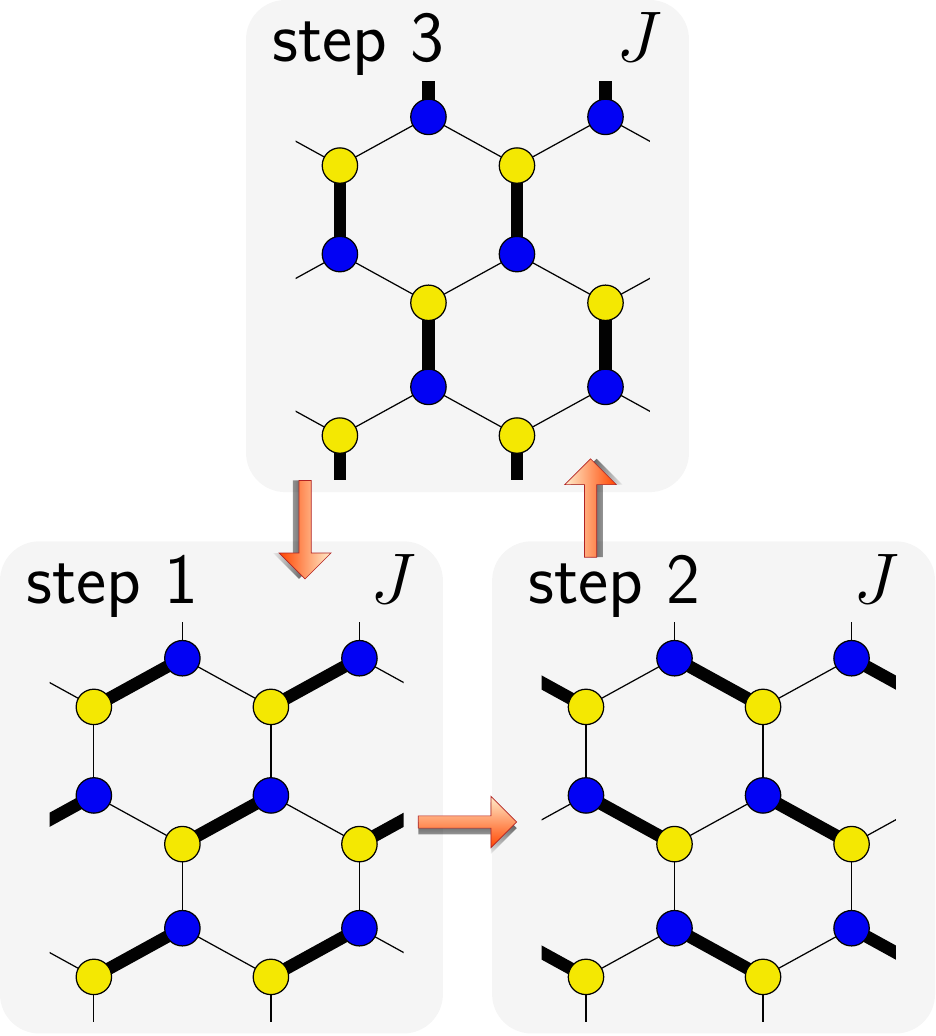}
\caption{Sketch of the three central steps of the driving protocol from Ref.~\cite{KitagawaPRB} acting on a hexagonal lattice.}
\label{app:fig:Prot3}
\end{figure}
\begin{figure}
\includegraphics[width=\columnwidth]{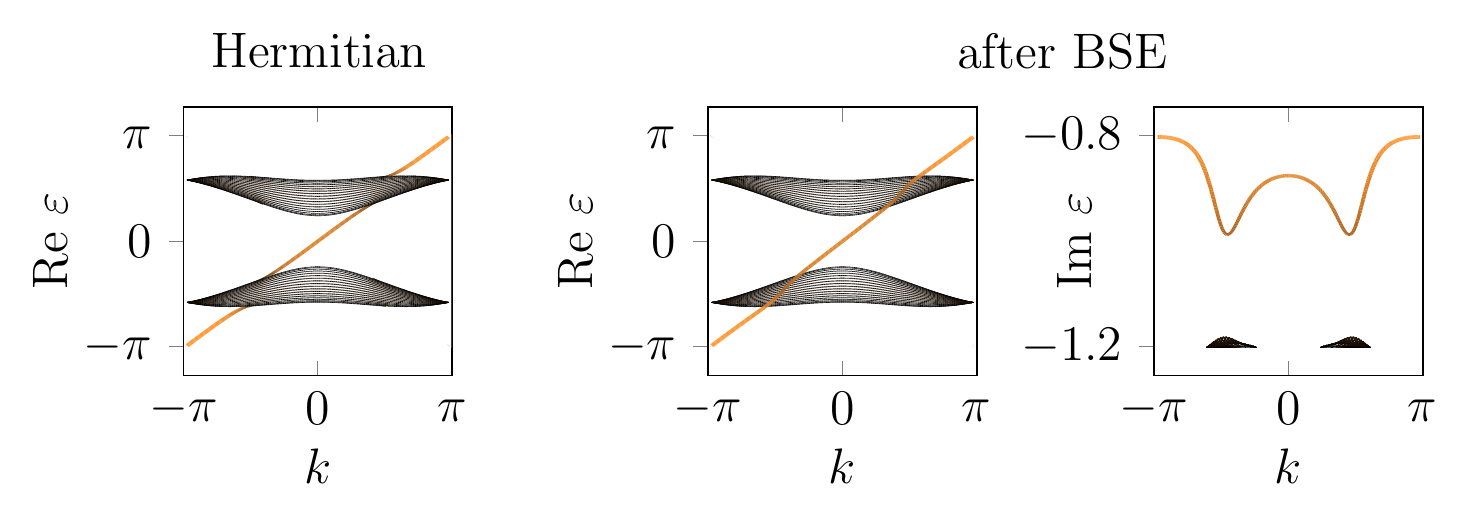}
\caption{
Left panel: Anomalous Floquet topological phase in the driving protocol from Ref.~\cite{KitagawaPRB} (with coupling $J=1.3$).
Central and right panel: 
Bulk and boundary states after BSE (with bulk losses $\gamma = 0.4$).
}
\label{app:fig:Prot4}
\end{figure}

As the second example, we consider the related driving protocol from Ref.~\cite{KitagawaPRB},
which takes place on a hexagonal instead of square lattice (see Fig.~\ref{app:fig:Prot3}).
BSE proceeds exactly as before, assigning uniform losses to the bulk ($\gamma$) but not the boundary sites,
and allows for complete detachment of boundary from bulk states, or for any other suitable manipulation (see Fig.~\ref{app:fig:Prot4}).

\subsection{Driven Kane-Mele model}

\begin{figure}
\includegraphics[width=\columnwidth]{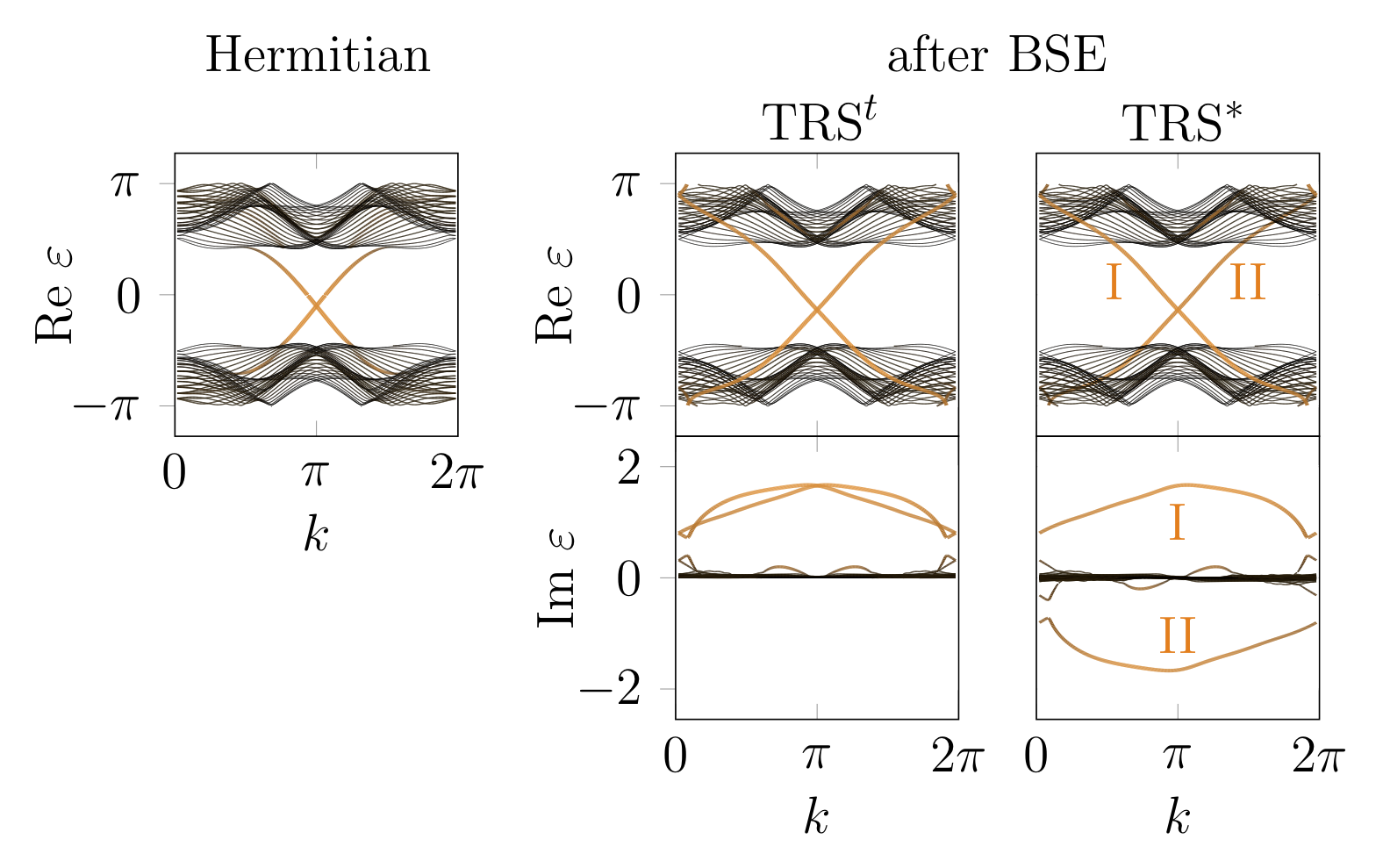}
\caption{Left panel: Anomalous Floquet topological $\mathbb Z_2$-phase
in the driven Kane-Mele model, with counterpropagating states on a zigzag boundary.
Model parameters are $J_1=J_2=1$, $\lambda_{\nu}=0.3$, $\lambda_{\mathrm{SO},1}=0.25$, $\lambda_{\mathrm{SO},2}=0.5$, $T=2\pi/\omega = 1.5$.
Central and right panel: Bulk and boundary states after BSE, with \TRST{} and \TRSC{} symmetry (for $\gamma_{t,*}=-1.25$, see text).
}
\label{app:fig:KM}
\end{figure}

The third example is a periodically driven Kane-Mele model, which has been considered, e.g., in the context of cold atomic systems~\cite{Yan2015}. This example does not belong to the category of driving protocols with discrete steps,
but features a continuous time dependence. The Hamiltonian
\begin{equation}\label{app:eq:KM}
\begin{aligned}
H_{\mathrm{KM}}(t)=&(J_1+J_2\cos(\omega t))\sum_{\langle i j\rangle} c_i^{\dagger} c_j +\lambda_{\nu} \sum_i \xi_i c_i^{\dagger} c_i\\&+\ii (\lambda_{\mathrm{SO},1}+\lambda_{\mathrm{SO},2} \cos(\omega t)) \sum_{\langle \langle ij \rangle \rangle} \nu_{ij} c_i^{\dagger} \sigma_z c_j
\end{aligned}
\end{equation}
extends the celebrated Kane-Mele model~\cite{KaneMelePRL} with a periodic modulation of the hopping ($\propto\! J_2)$ and spin-orbit coupling ($\propto \lambda_{\mathrm{SO},2})$.
The driven Kane-Mele model still possesses fermionic TRS, and supports an anomalous $\mathbb Z_2$ topological Floquet phase (left panel in Fig.~\ref{app:fig:KM}).
The anomalous nature of this phase is not obvious from the quasienergy plot in Fig.~\ref{app:fig:KM}, but is corroborated by the vanishing $\mathbb Z_2$ invariant of the Floquet bands.

\begin{figure}
\includegraphics[width=0.8\columnwidth]{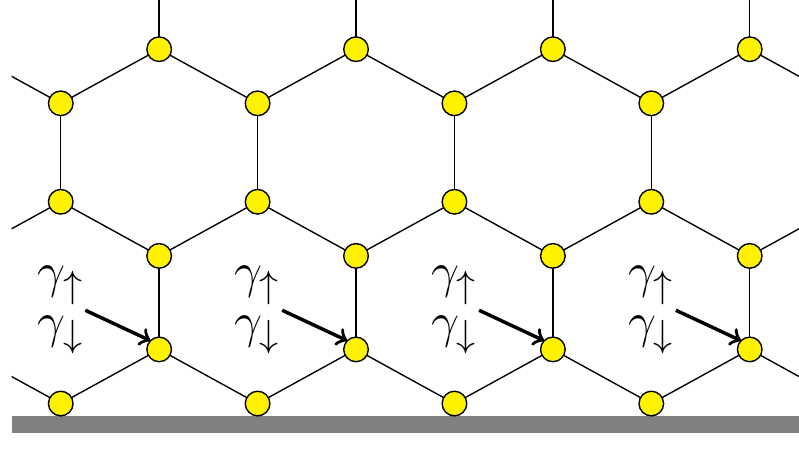}
\caption{Assignment of boundary losses for BSE in the driven Kane-Mele model, for a zigzag boundary of the hexagonal lattice.
Spin-dependent losses $\gamma_\uparrow$, $\gamma_\downarrow$ are introduced on every second boundary site, as indicated by the arrows.}
\label{app:fig:hexa_latt}
\end{figure}

To apply BSE we introduce boundary losses, similar to the examples of the driving protocols.
For a zigzag boundary, a minimal assignment is shown in Fig.~\ref{app:fig:hexa_latt},
selecting sites with a large amplitude of the boundary state wave function.
The boundary losses $\gamma_\uparrow$, $\gamma_\downarrow$ depend on the spin degree of freedom in the Kane-Mele model. To preserve TRS, it must $\gamma_{\uparrow}=\gamma_{\downarrow}=\gamma_t$ for \TRST{} and
$\gamma_{\uparrow}=-\gamma_{\downarrow}=\gamma_*$ for \TRSC{}.

The central and right panel in Fig.~\ref{app:fig:KM} show how BSE allows us to detach the boundary states from the bulk bands, in exact analogy to the result for the driving protocols. The topological preservation and the consequences of TRS could now be discussed along the lines established in the main text for the TR symmetric protocol.
Note that after BSE, the real part of the quasienergy dispersion of the boundary states visibly crosses the bulk bands and connects at the $\pm \pi$ quasienergy (which nicely shows the anomalous nature of this phase).
This apparent crossing, which we also observe in Fig.~\ref{fig:dispersion} in the main text, becomes possible since BSE separates the imaginary part of the boundary states and bulk bands.
In the complex plane (recall Fig.~\ref{fig:concept} in the main text), the quasienergy dispersions are fully separated and do not cross.

